\documentclass{article}
\usepackage{spconf,amsmath,graphicx}

\usepackage{enumitem}
\usepackage{color}
\usepackage{pifont}
\setlist{nosep, leftmargin=14pt}

\usepackage{mwe} % to get dummy images

\newcommand{\cmark}{\ding{51}}%
\newcommand{\xmark}{\ding{55}}%
\title{Liver Fibrosis and NAS scoring from CT images using self-supervised learning and texture encoding}
 \name{Ananya Jana$^{\star}$, Hui Qu$^{\star}$,  Carlos D. Minacapelli$^{\dagger}$,  Carolyn Catalano$^{\dagger}$, Vinod Rustgi$^{\dagger}$,  Dimitris Metaxas$^{\star}$}
 \address{$^{\star}$ Department of Computer Science, Rutgers University\\  $^{\dagger}$Rutgers Robert Wood Johnson Medical School, Division of
Gastroenterology and Hepatology}
\begin{document}
%\ninept
%
\maketitle
\begin{abstract}
Non-alcoholic fatty liver disease (NAFLD) is one of the most common causes of chronic liver diseases (CLD) which can progress to liver cancer. The severity and treatment of NAFLD is determined by NAFLD Activity Scores (NAS) and liver fibrosis stage, which are usually obtained from liver biopsy. However, biopsy is invasive in nature and involves risk of procedural complications. 
Current methods to predict the fibrosis and NAS scores from noninvasive CT images rely heavily on either a large annotated dataset or transfer learning using pretrained networks. However, the availability of a large annotated dataset cannot be always ensured and there can be domain shifts when using transfer learning. In this work, we propose a self-supervised learning method to address both problems. As the NAFLD causes changes in the liver texture, we also propose to use texture encoded inputs to improve the performance of the model.
Given a relatively small dataset with 30 patients, we employ a self-supervised network which achieves better performance than a network trained via transfer learning. The code is publicly available\footnote{https://github.com/ananyajana/fibrosis{\_}code}.
\end{abstract}
\section{Introduction}
\label{sec:intro}
NAFLD is growing in prevalence in the worldwide population~\cite{asrani2019burden}. If not diagnosed timely, it can gradually progress to liver cancer~\cite{bedossa2017pathology}. The diagnosis of NAFLD requires histo-pathological examination of the fibrosis and NAS scores. The NAS scores comprise of three different scores: steatosis, lobular inflammation and ballooning. The gold standard in determining these scores is liver biopsy which is an invasive procedure and involves risks such as bleeding, infections, etc. Thus it would be beneficial if non-invasive imaging modalities can be used to correctly stage fibrosis and NAS.
In non-invasive imaging modalities, although MRI contains more information than CT, the latter is cheaper and hence it is of great value to detect fibrosis and NAS scores from CT images.

There are works that stage liver fibrosis from CT using deep learning based techniques~\cite{yasaka2018deep}. Choi et al.~\cite{choi2018development} train convolutional neural networks from scratch using a dataset of over 4,000 subjects. Such a large amount of fibrosis and NAS labeled data might not be always available as the labeling requires the invasive biopsy procedure. To train a good model on a small dataset, transfer learning is usually adopted to make use of the pretrained weights from large datasets. For example, in~\cite{jana2020deep} the authors stage NAS and fibrosis by initializing a ResNet-18~\cite{he2016deep} model with pretrained weights from ImageNet dataset.
Such type of pretraining not only requires a huge amount of labeled data, which is often infeasible in the medical image field, but also suffers from  domain shift between natural images in ImageNet dataset and CT images in the small medical dataset.

In this work we address these limitations by using self-supervised learning. Self-supervised learning (SSL) uses a pretext task to pretrain the network with unlabeled data. The pretrained network is then finetuned in the downstream tasks. An advantage of SSL is that we can make use of a large amount of unlabeled data to train the network. This bypasses the need for a huge dataset with fibrosis and NAS labels for our classification task.
Besides, the problem of domain shift is also effectively addressed as both the self-supervised pretraining and downstream classification are on medical images.

In SSL, it is important to select a suitable pretext task to make the pretraining effective. The pretext task should help the network understand the features that are needed to solve the downstream task. We choose the context restoration method~\cite{chen2019self} as our pretext task. In this method, the network learns by reconstructing images from their corrupt versions using the context or neighborhood pixels information. This knowledge makes this self-supervised training a fit candidate for fibrosis and NAS score classification task because these diseases can bring in granular textural changes in liver.

\begin{figure*}[t]
\centering
\includegraphics[width=0.85\textwidth]{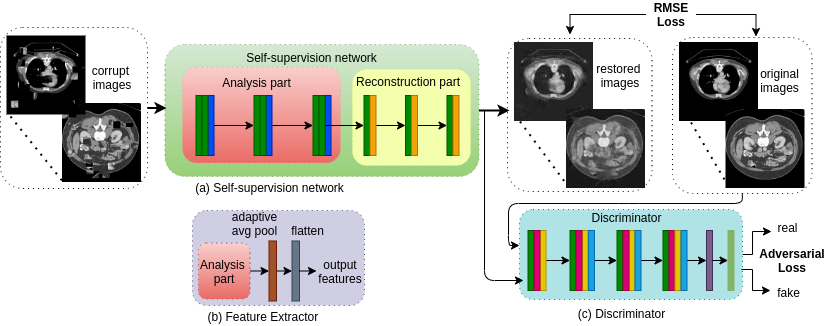}
\caption{(a) Self-supervised learning using the context restoration pretext. The self-supervision network consists of repeated convolution (green), pooling (blue) and upsampling (orange) layers. The analysis part is fine-tuned in the downstream task. RMSE loss and adversarial loss are used to train the network. (b) Feature extractor constructed by using the analysis part followed by adaptive average pooling.(c) A discriminator is used for adversarial loss. Different colors are convolution (green), leaky-relu (red), dropout (yellow), batch normalization (blue), fc (violet) and sigmoid (light green) layers.} 
\label{fig:ssl}
\end{figure*}

With the progression of NAFLD, the liver may undergo changes in texture or liver nodularity~\cite{kim2019development,lubner2017texture,lubner2018accuracy,lubner2019ct}, in the ratio of the volumes of the liver's left and right lobes~\cite{hunt2016liver}, in contour~\cite{kim2019development}, and expansion of the gall bladder fossa region~\cite{mamone2019expanded} etc.  To assist the network in better capturing the visual changes in the liver texture, we use texture encoded input~\cite{fetit2020training} instead of the original CT images when fine-tuning the network in the downstream classification task.
In summary, our main contributions are:
\begin{itemize}
    \item Introduction of an effective self-supervision approach for fibrosis and NAS score prediction.
    \item Demonstration that texture encoded input (local binary pattern) can be effectively utilized to achieve good performance for fibrosis and NAS classification.
\end{itemize}

\section{Methods}

Our proposed method has three steps: (1) Data preprocessing, (2) self-supervised pretraining using context restoration, and (3) fibrosis stage  and NAS scores classification by fine-tuning the SSL pretrained network with local binary pattern inputs.

\vspace{-0.2cm}
\subsection{Data Preprocessing}
The 3D CT volume of a subject consists of raw CT slices that are not suitable to be directly input into a neural network. We first performed Hounsfield windowing on these slices with the range $[-200, 250]$ to improve the image contrast and then segmented out the liver part using a U-net~\cite{ronneberger2015u} like in~\cite{jana2020deep} to get rid of irrelevant organs in the CT images.
After liver segmentation in the 2D slices from all subjects, we discarded the slices which had either no liver present or very small liver portion (by using mean pixel value 5 as threshold). The remaining 2D slices were resized to $224\times224$ resolution.

\begin{figure*}[t]
\centering
\includegraphics[width=0.9\textwidth]{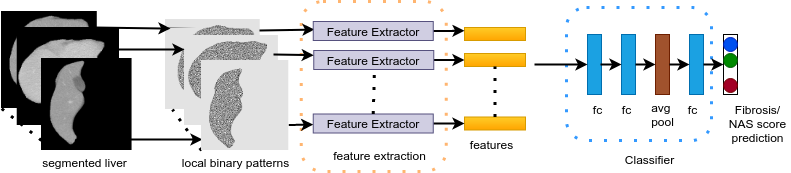}
\caption{Fibrosis \& NAS score classification: The images are converted to their texture encoded representation and the feature extractor generates the features from these texture encoded input. The final classifier uses these features to predict a final score}
\label{fig:classifier}
\end{figure*}

\begin{table*}[t]
    \centering
    \caption{Original and combined Fibrosis stage and NAS score distributions in the 30-subject dataset.}
    \begin{tabular}{|l|p{0.4cm}|p{0.4cm}|p{0.4cm}|p{0.4cm}|p{0.4cm}|p{0.4cm}||p{0.4cm}|p{0.4cm}|p{0.4cm}|p{0.4cm}||p{0.4cm}|p{0.4cm}|p{0.4cm}|p{0.4cm}||p{0.7cm}|p{0.7cm}|p{0.7cm}|}
    \hline
         & \multicolumn{6}{|c||}{Fibrosis} & \multicolumn{4}{|c||}{NAS steatosis} & \multicolumn{4}{|c||}{NAS lobular}  & \multicolumn{3}{|c|}{NAS ballooning}\\\hline
    Score & 0 & 1 & 2 & 3 & 3.5 & 4 &  0 & 1 & 2 & 3 & 0 & 1 & 2 &3 &0 &1 &2\\ \hline
    Original & 7 & 6 & 4 & 3 & 2 & 8 & 2& 9 & 11 & 8 & 9 & 10 & 8 & 3 & 8 & 11 & 11 \\ \hline
    Combined & 7 & \multicolumn{2}{c|}{10} & \multicolumn{3}{c||}{13} & 
          \multicolumn{2}{|c|}{11} & \multicolumn{2}{c||}{19} &  9 & 10 & \multicolumn{2}{|c||}{11} & 8 & 11 & 11 \\ \hline
    \end{tabular}
    \label{tab:score}
\end{table*}

\vspace{-0.2cm}
\subsection{Self-Supervised Pretraining by Context Restoration}
As texture change in the liver is one of the key indicators of NAFLD, it becomes extremely important that the network understands the context of a pixel in conjunction with its neighborhood rather than in isolation. Therefore, we choose context restoration~\cite{chen2019self} as the pretext task. This task aims to reconstruct the original image from the corrupted version. The original image is corrupted by swapping a pair of non-overlapping random patches in the image. This step is repeated several times to generate the final corrupted image. The difficulty level of this restoration task increases as the number of swaps increases or if the patch size increases.
In the restoration process the network can learn important semantic features by focusing on the neighborhood of a patch.

Our self-supervised pretraining step is shown in Fig.~\ref{fig:ssl}(a). The network consists of an analysis/encoder part and a reconstruction/decoder part. The analysis part has three convolution and max-pooling layers and is used for feature extraction. The reconstruction part consists of repeated convolution and upsampling layers and aims to reconstruct the original image from features.
% The analysis part consists of repeated convolutions followed by max-pooling layers. The reconstruction part consists of repeated convolution and upsampling layers. The analysis  part is used for feature extraction. 
In the downstream classification task, the analysis part followed by an adaptive average pooling layer (Fig.~\ref{fig:ssl}(b)) is used as the feature extractor to obtain a feature vector for each slice.
% The feature extractor also contain an adaptive average pooling layer which collects the global features from the local features of an individual 2d slices. %The kernels are same for all the convolution layers with size $3x3$, stride=$1$ and padding=$1$. 
Chen et al.\cite{chen2019self} used MSE loss for training the task. We choose RMSE over MSE to prevent the possible outliers from perturbing the network drastically. Besides, the changes brought in by NAFLD could be very subtle, almost imperceptible to ordinary vision. Hence we introduce an adversarial loss to help the network learn to restore the context in a more realistic way. The discriminator consists of convolution layers, leaky-relu layers, dropout and batch normalization layers followed by a final sigmoid layer.

\vspace{-0.2cm}
\subsection{Local Binary Pattern} NAFLD can bring in a number of morphological changes in the liver - change in liver texture being one of the key changes.
There are studies~\cite{kim2019development} that explore the manual scoring of the liver texture change by using quantification measures like Liver Surface Nodularity (LSN) score. In this work we explore if our network can leverage the information contained in texture encoded input to classify the disease better. 
We use a computationally simple and effective method called local binary pattern. In this method each pixel is encoded in a way such that its surrounding pixels information is included in it. Suppose there are $n$ neighbor pixels, then the central pixel can be thought of as an $n$ bit representation where each bit represents a particular neighbor pixel's relative intensity value. When a neighbor pixel's intensity value is greater than the central pixel, the corresponding bit position contains a 1, otherwise 0. The formulation is written as~\cite{pietikainen2010local}.
\begin{equation}
LBP(x_c, y_c) = \sum_{p=0}^{p=n-1}{f(i_p - i_c)}2^p
\end{equation}
where $f(x)=1$ if $x\geq 1$, and 0 otherwise, $(x_c, y_c)$ is the central pixel with intensity $i_c$, and $i_p$ denotes the intensity of the neighboring pixel. We compute the local binary pattern by taking a radius of 1 and an 8-connected neighborhood. We choose radius 1 because the changes brought by liver fibrosis are very granular, almost imperceptible to human eyes and hence the smallest possible radius may work best to capturing the pixel differences.

The local binary patterns of the images are used as input to the classification network. The self-supervision part is trained entirely on CT images, not the local binary patterns.

\subsection{Classification}
The network architecture of the fibrosis and NAS score classification is shown in Fig.~\ref{fig:classifier}. It consists of the feature extractor (Fig.~\ref{fig:ssl}(b)) and a final classifier. The feature extractor consists of the analysis part of the self-supervision network and an adaptive average pooling layer and aims to extract local features for each individual 2D slice. Because we want to predict the patient-wise scores, the first two fully-connected (fc) layers process the local features and then the average pooling layer obtains the global feature for each patient from local features. And this global feature is used for prediction through the last fc layer. The cross entropy loss is adopted to train the classification network.

\begin{table*}[t]
\centering
\caption {Mean AUC values of fibrosis and NAS scores prediction using different methods (Three-fold cross validation, average of 5 repeated experiments).}
\begin{tabular}{|c |l |l |l |l |}
\hline
Method           & Fibrosis & NAS steatosis & NAS lobular  & NAS ballooning  \\ \hline

Pretrained Resnet~\cite{jana2020deep} on CT  & 76.35$\pm$15.77 &	67.88$\pm$8.85 &	61.82$\pm$14.19	& 64.38$\pm$16.29 \\

Simulation of Choi et al.~\cite{choi2018development} & 69.26$\pm$4.7	& - &	- &	-  \\

Ours (SSL+LBP+Adversarial Loss) & \textbf{78.47$\pm$1.97}	& \textbf{84.66$\pm$13.27} & \textbf{74.44$\pm$1.71}& \textbf{73.90$\pm$2.86}\\
\hline
\end{tabular}
\label{tab:combined}
\end{table*}

\begin{table*}[t]
\centering
\caption {Ablation Study. Best results shown in bold and second best results in underline.}
\begin{tabular}{|c |c|c|c|l |l |l |l |}
\hline
Method  & SSL & Adv. Loss & image/LBP &  Fibrosis & NAS steatosis & NAS lobular  & NAS ballooning  \\ \hline
Ablation1 & \xmark & \xmark & image  & 58.61$\pm$5.82& 75.16$\pm$14.13 & 68.48$\pm$4.5 & 61.48$\pm$1.41\\
 
Ablation2 & \xmark & \xmark & LBP &  58.82$\pm$9.57& 48.38$\pm$13.36 & 47.57$\pm$7.49 & 47.17$\pm$6.46\\

Ablation3 & \cmark & \xmark & image  & 77.97$\pm$3.16& 76.36$\pm$9.89 & 69.33$\pm$2.27 & 60.9$\pm$5.94\\

Ablation4 & \cmark &\xmark & LBP &  72.51$\pm$4.13& 76.59$\pm$16.15 & 73.88$\pm$1.39 & 71.23$\pm$3.21\\

Ablation5 & \cmark & \cmark & image &\textbf{79.30$\pm$1.57}& \underline{78.96$\pm$7.6} & \underline{73.44$\pm$1.36} & \underline{65.59$\pm$2.78}\\
Ours &\cmark & \cmark & LBP &  \underline{78.47$\pm$1.97}	& \textbf{84.66$\pm$13.27} & \textbf{74.44$\pm$1.71}& \textbf{73.9$\pm$2.86}\\

\hline
\end{tabular}
\label{tab:ablation}
\end{table*}

\section{Experiments}
\vspace{-0.2cm}
\subsection{Dataset and Evaluation Metrics}

The pretext task of self-supervised learning is trained on the MICCAI 2017 LiTS challenge dataset~\cite{bilic2019liver}. It is a large dataset with high image quality. Most importantly, the liver images are similar as those in our dataset for fibrosis and NAS scores prediction. LiTS dataset has a total of 201 patients CT volumes, and there are 131 and 70 cases in the training and test sets, respectively. The segmentation labels in LiTS dataset are not used during self-supervised learning.

Our dataset for fibrosis and NAS score classification is the same as~\cite{jana2020deep}. The dataset comprises of 30 patients CT volumes. The distribution of the fibrosis and NAS scores are shown in Table~\ref{tab:score}. In the original data, the number of patients in some of the classes was very small and hence some of the classes have been combined(`Combined' row) after discussion with our collaborating doctors.

We use the Area Under ROC Curve (AUC) to evaluate the performance of our model.

\vspace{-0.2cm}
\subsubsection{Implementation Details} 
The self-supervision network was trained using the Adam optimizer for 700 epochs with batch size 30 and learning rate 0.0002. The images were normalized using a mean 0.485 and std 0.229. We used a patch size of 20 pixels and a total of 10 iterations to corrupt the images. The adversarial loss used was the binary cross entropy loss. 
In the downstream classification task, we freeze all except the last two convolution layers of the analysis part of the self supervision network. This can avoid overfitting and potential memory problems.

The model was trained using the Adam optimizer for 30 epochs with a learning rate 0.0001,  batch size 4 and weight decay 0.01. The kernels are the same for all the convolution layers with size $3x3$, stride=$1$ and padding=$1$. We select the best performing model of the 30 epochs for test.

\vspace{-0.2cm}
\subsection{Results and Discussion}
\subsubsection{Comparison with state-of-the-art methods}
We compare our method with two state-of-the-art methods~\cite{choi2018development,2009.10687}. In~\cite{choi2018development}, the model was trained using a large annotated dataset. The network in~\cite{choi2018development} is a 3D network, and we made a 2D simulation of it for comparison. They used a network of 15 convolutions and pooling layers and a final average pooling layer. As our image size was 224x224 and we didn't use any upsampling layers, we used a network comprising of 15 convolutions and 5 max pooling layers and a final average pooling layer. In~\cite{jana2020deep}, the authors utilized transfer learning to the classification model. The ResNet-18 based feature extractor was initialized with ImageNet pretrained weights.
% We simulate the work~\cite{2009.10687} on lbp as well. 

The results are shown in Table~\ref{tab:combined}.
Our results outperform both methods in all tasks. Compared to Choi et. al~\cite{choi2018development}, our method is more focused on learning specific features of the disease(i.e. texture) and hence achieves better performance.
Compared to the transfer learning method~\cite{jana2020deep}, our method takes advantages of the self-supervised learning and has no domain shift problems, therefore achieves much better performance. Another advantage of our method is that our network is more suitable for this small patient dataset as it is more lightweight and is not prone to overfit the data compared with the other two methods. Our self-supervision focuses specifically on understanding the pixel neighborhood information better which is very important for fibrosis and NAS score classification. The other two previous works already mentioned are more generic in nature and hence may not leverage the disease specific features to the best possible extent.

\vspace{-0.2cm}
\subsubsection{Ablation Study}
We performed ablation studies to illustrate the effectiveness of the proposed method. The results are shown in Table~\ref{tab:ablation}. It can be observed that the self-supervised pretraining, the adversarial loss and the local binary pattern inputs all have positive effects on the classification performance.
Without the self-supervised pretraining, the results (Ablation1 and Ablation2) are much worse compared with those using SSL, because it is difficult for the network to learn meaningful features on the small dataset. The self-supervised pretraining provides good initial weights for the classification task. The adversarial loss in the pretext task can help the network learn to restore the context in a more realistic way, thus providing a better pretrained feature extractor for the downstream task. As for the local binary pattern, it is not useful when training from scratch, probably because of the small size of dataset. With the help of SSL, the LBP are better compared to using the original images as input.

\section{Conclusion}
In this work, we proposed a self-supervised learning based method to predict liver fibrosis stage and NAS scores from CT images. The self-supervised pretraining can solve the problems of the scarcity of large annotated dataset and domain shift in transfer learning. Besides, we adopted texture encoded input (local binary pattern) as the network input instead of original images to better capture the texture changes. Our method outperformed the state-of-the-art methods. %Our method achieved much better performance compared with state-of-the-art methods.

\section{Compliance with Ethical Standards}
\label{sec:ethics}
This research study includes private data of human subjects from our collaborative partner and it was approved by the Rutgers University Institutional Review Board (IRB).

\section{Acknowledgments}
\label{sec:acknowledgments}
This research was funded based on partial funding to D. Metaxas from NSF: IIS-1703883, CNS-1747778, CCF-1733843, IIS-1763523, IIS-1849238 – 825536 and MURI-Z8424104 -440149.
%  \item ``This work was supported by […] (Grant numbers) and […]. Author X has served on %advisory boards for Company Y.''

\bibliographystyle{IEEEbib}
\bibliography{strings,refs}

\begin{thebibliography}{10}

\bibitem{asrani2019burden}
Sumeet~K Asrani, Harshad Devarbhavi, John Eaton, and Patrick~S Kamath,
\newblock ``Burden of liver diseases in the world,''
\newblock {\em Journal of hepatology}, vol. 70, no. 1, pp. 151--171, 2019.

\bibitem{bedossa2017pathology}
Pierre Bedossa,
\newblock ``Pathology of non-alcoholic fatty liver disease,''
\newblock {\em Liver International}, vol. 37, pp. 85--89, 2017.

\bibitem{yasaka2018deep}
Koichiro Yasaka, Hiroyuki Akai, Akira Kunimatsu, Osamu Abe, and Shigeru Kiryu,
\newblock ``Deep learning for staging liver fibrosis on ct: a pilot study,''
\newblock {\em European radiology}, vol. 28, no. 11, pp. 4578--4585, 2018.

\bibitem{choi2018development}
Kyu~Jin Choi, Jong~Keon Jang, Seung~Soo Lee, Yu~Sub Sung, Woo~Hyun Shim,
  Ho~Sung Kim, Jessica Yun, Jin-Young Choi, Yedaun Lee, Bo-Kyeong Kang, et~al.,
\newblock ``Development and validation of a deep learning system for staging
  liver fibrosis by using contrast agent--enhanced ct images in the liver,''
\newblock {\em Radiology}, vol. 289, no. 3, pp. 688--697, 2018.

\bibitem{jana2020deep}
Ananya Jana, Hui Qu, Puru Rattan, Carlos~D Minacapelli, Vinod Rustgi, and
  Dimitris Metaxas,
\newblock ``Deep learning based nas score and fibrosis stage prediction from ct
  and pathology data,''
\newblock in {\em 2020 IEEE 20th International Conference on Bioinformatics and
  Bioengineering (BIBE)}. IEEE, 2020, pp. 981--986.

\bibitem{he2016deep}
Kaiming He, Xiangyu Zhang, Shaoqing Ren, and Jian Sun,
\newblock ``Deep residual learning for image recognition,''
\newblock in {\em Proceedings of the IEEE conference on computer vision and
  pattern recognition}, 2016, pp. 770--778.

\bibitem{chen2019self}
Liang Chen, Paul Bentley, Kensaku Mori, Kazunari Misawa, Michitaka Fujiwara,
  and Daniel Rueckert,
\newblock ``Self-supervised learning for medical image analysis using image
  context restoration,''
\newblock {\em Medical image analysis}, vol. 58, pp. 101539, 2019.

\bibitem{kim2019development}
Tae-Hoon Kim, Ji~Eon Kim, Jong-Hyun Ryu, and Chang-Won Jeong,
\newblock ``Development of liver surface nodularity quantification program and
  its clinical application in nonalcoholic fatty liver disease,''
\newblock {\em Scientific reports}, vol. 9, no. 1, pp. 1--10, 2019.

\bibitem{lubner2017texture}
Meghan~G Lubner, Kyle Malecki, John Kloke, Balaji Ganeshan, and Perry~J
  Pickhardt,
\newblock ``Texture analysis of the liver at mdct for assessing hepatic
  fibrosis,''
\newblock {\em Abdominal Radiology}, vol. 42, no. 8, pp. 2069--2078, 2017.

\bibitem{lubner2018accuracy}
Meghan~G Lubner, Daniel Jones, Adnan Said, John Kloke, Scott Lee, and Perry~J
  Pickhardt,
\newblock ``Accuracy of liver surface nodularity quantification on mdct for
  staging hepatic fibrosis in patients with hepatitis c virus,''
\newblock {\em Abdominal Radiology}, vol. 43, no. 11, pp. 2980--2986, 2018.

\bibitem{lubner2019ct}
Meghan~G Lubner, Daniel Jones, John Kloke, Adnan Said, and Perry~J Pickhardt,
\newblock ``Ct texture analysis of the liver for assessing hepatic fibrosis in
  patients with hepatitis c virus,''
\newblock {\em The British journal of radiology}, vol. 92, no. 1093, pp.
  20180153, 2019.

\bibitem{hunt2016liver}
Oliver~F Hunt, Meghan~G Lubner, Timothy~J Ziemlewicz, Alejandro~Mu{\~n}oz del
  Rio, and Perry~J Pickhardt,
\newblock ``The liver segmental volume ratio (lsvr) for non-invasive detection
  of cirrhosis: Comparison with established linear and volumetric measures,''
\newblock {\em Journal of computer assisted tomography}, vol. 40, no. 3, pp.
  478, 2016.

\bibitem{mamone2019expanded}
Giuseppe Mamone and Roberto Miraglia,
\newblock ``The “expanded gallbladder fossa sign” in liver cirrhosis,''
\newblock {\em Abdominal Radiology}, vol. 44, no. 3, pp. 1199--1200, 2019.

\bibitem{fetit2020training}
Ahmed~E Fetit, John Cupitt, Turkay Kart, and Daniel Rueckert,
\newblock ``Training deep segmentation networks on texture-encoded input:
  application to neuroimaging of the developing neonatal brain,''
\newblock in {\em Medical Imaging with Deep Learning}, 2020.

\bibitem{ronneberger2015u}
Olaf Ronneberger, Philipp Fischer, and Thomas Brox,
\newblock ``U-net: Convolutional networks for biomedical image segmentation,''
\newblock in {\em International Conference on Medical image computing and
  computer-assisted intervention}. Springer, 2015, pp. 234--241.

\bibitem{pietikainen2010local}
Matti Pietik{\"a}inen,
\newblock ``Local binary patterns,''
\newblock {\em Scholarpedia}, vol. 5, no. 3, pp. 9775, 2010.

\bibitem{bilic2019liver}
Patrick Bilic, Patrick~Ferdinand Christ, Eugene Vorontsov, Grzegorz Chlebus,
  Hao Chen, Qi~Dou, Chi-Wing Fu, Xiao Han, Pheng-Ann Heng, J{\"u}rgen Hesser,
  et~al.,
\newblock ``The liver tumor segmentation benchmark (lits),''
\newblock {\em arXiv preprint arXiv:1901.04056}, 2019.

\end{thebibliography}

\end{document}